\documentclass[12pt]{article}
\usepackage{epsf}

\newcommand{\beq}{\begin{equation}}
\newcommand{\eeq}{\end{equation}}
\newcommand{\beqcol}{\begin{array}{rcl}}
\newcommand{\eeqcol}{\end{array}}

\newcommand{\eps}{\varepsilon}

\newcommand{\nin}{\in{\hspace{-3.5mm}}/}

\def\res{\mathop{\mbox{\rm{res}}}}

\begin{document}

{\thispagestyle{empty}

\begin{flushright}  
KCL-MTH-99-28 \\
July 1999 
\end{flushright}
\vfill
\begin{center}
{\Large 
Replica-deformation of the $SU(2)$-invariant Thirring model  \\[3mm]
via solutions of the qKZ equation}
\\
\vspace{2cm}
{\sc Mathias Pillin}   \\
\vspace{1cm}
Department of Mathematics, King's College \\
Strand, London WC2R 2LS, U.K.   \\
e-mail: map@mth.kcl.ac.uk 
\medskip

\bigskip

\begin{center}
{\it Dedicated to the memory of Mosh\'e Flato}
\end{center}

\vfill
{\bf Abstract}
\end{center}
\begin{quote}
The response of an integrable QFT under variation of the Unruh 
temperature has recently been shown to be computable from an 
$S$-matrix preserving (``replica'') deformation of the form factor approach. 
We show that replica-deformed form factors of the 
$SU(2)$-invariant Thirring model can be found among the 
solutions of the rational $sl_2$-type quantum Knizhnik-Zamolodchikov 
equation at generic level. We show that modulo conserved charge solutions
the deformed form factors are in one-to-one correspondence to 
the ones at level zero and use this to conjecture the deformed form 
factors of the Noether current in our model.
\end{quote}
\eject
}

\setcounter{page}{1}

{\section{Introduction}}

\bigskip
\bigskip

The form factor approach to massive integrable quantum field 
theories (QFTs) \cite{KW,SMIR1} provides a powerful tool to compute 
exact matrix elements of local operators from the knowledge
of the (factorized) S-matrix. To be precise, a form factor 
in this context is the matrix element of some local operator between 
the physical vacuum and a multi-particle scattering state.
These form factors contain all the information 
about the QFT as the Wightman functions can in principle 
be (re-)constructed in terms of them by saturating with
a complete set of scattering states. Alternatively each 
local operator can be set into correspondence to a 
sequence of form factors; a feature that has
been used to classify the full content of local operators in 
certain models, see e.g.~\cite{SMIR3,MAP2}. 

For an integrable QFT 
the form factors can be characterized axiomatically as 
tensor-valued meromorphic functions obeying a recursive set 
of functional equations, the so called form factor equations \cite{SMIR1}.

As this will be of importance later we anticipate that one of these 
equations (the cyclic form factor equation), see equation 
(\ref{watson}) below, is closely 
related to the Bisignano-Wich\-mann-Unruh thermalization phenomenon 
\cite{BW,UN}. In particular the cyclicity parameter $\beta = 2\pi $ 
is both physically and mathematically linked to the Unruh 
temperature $T_U = 1/2\pi$ (in natural units), seemingly of a very 
different origin. This connection that has been used to formulate the 
cyclic form factor equation without 
reference to bootstrap analyticity properties and to generalize it 
to non-integrable QFTs \cite{MAX3}.   

Among the most prominent non-scalar massive integrable models 
which have been solved by the form factor method at $T_U= 1/ 2\pi$ 
are the Sine-Gordon model and the $SU(2)$-invariant Thirring model 
\cite{SMIR1,SMIR2}. We will refer to the latter, which will 
concern us in this paper, 
simply as the Thirring model (even though this clashes a bit 
with standard usage).

\medskip

On the other hand, in an a priori independent direction of 
research it was shown in a pioneering work \cite{FR} that 
the classical Knizhnik-Zamolodchikov differential 
equation (for a survey 
see \cite{EFK}) which arises in conformal field theory 
admits a quantum deformation. This deformation amounts 
in a consistent system of holonomic difference 
equation for a tensor valued meromorphic function, 
which is referred to as deformed or quantum 
Knizhnik-Zamolodchikov (qKZ) equation. The authors 
of \cite{FR} already realised that a subset of the 
standard form factor equations \cite{SMIR1} provide 
a very specific example of a qKZ equation. Much work 
has been done during recent years, to elaborate this 
connection, see \cite{JM1,NPT,MT} and references therein.

\medskip

In the seminal paper \cite{TV} the complete set of solutions 
of the qKZ equation for a rational $sl_2$-type R-matrix has been found. 
In this case one considers a function $\Psi(z_1, \ldots , z_n)$ taking values 
in a tensor product of $sl_2$-modules $V_1\otimes \ldots \otimes V_n$, 
and subject to a system of difference equations with shift parameter $p$ 

\beq
\beqcol
\Psi(z_1, \ldots , z_m + p, \ldots , z_n ) & = &  
R_{m,m-1}(z_m-z_{m-1}+p ) \cdots R_{m,m-1}(z_m-z_{1}+p ) \times \\
& & {\hspace{-2cm}} \times 
R_{m,m-1}(z_m-z_{n} ) \cdots R_{m,m+1}(z_m-z_{1} ) 
\Psi(z_1, \ldots , z_m , \ldots , z_n ).
\eeqcol
\label{qKZ-def}
\eeq

Here, $R_{i,j}(z) \in {\rm End} (V_i\otimes V_j)$ is the 
rational $sl_2$ R-matrix. The more general case of this 
equation involving an extra parameter $\kappa$ as 
studied in \cite{TV} will not be of importance in this paper.

In a nut-shell the solutions of (\ref{qKZ-def}) are found as 
generalised hypergeometric integrals. Subsequently, the construction of 
\cite{TV} has been applied to irreducible representations of $sl_2$ in 
\cite{MV}. It is this case which is relevant here and which will 
provide our main mathematical tool.

\medskip

Returning to form factors, it has been shown in \cite{NPT} 
that Smirnov's solutions of the form factor equations in 
the Thirring model \cite{SMIR1,SMIR2} can indeed be extracted 
from the general solutions of (\ref{qKZ-def}). In particular, 
Smirnov's results follow if we take only 
two-dimensional fundamental representations $V$ of $sl_2$ 
(obviously corresponding to the physical states in the model) 
and set $ p =- 2 \pi i $. This particular choice is 
referred to as level zero solution of the qKZ equation. 
In other words, the form factors of the Thirring model 
can be identified with specific solutions of the qKZ 
equation at level zero.

\medskip

The physics problem we wish to solve is to construct the 
``replica-deformed''form factors in the sense of \cite{MAX1} 
for the SU(2)-invariant massive Thirring model.

We shall review some aspects of the replica-deformation 
in section 2. Here it may be sufficient to remark that in view of 
the aforementioned relation to the Bisignano-Wichmann-Unruh thermalization,
taking $p \neq 2\pi i$ in the qKZ equations amounts to studying 
the ``continuation'' of the original QFT to some ``off-critical'' Unruh 
temperature. The quest for such a continuation naturally arises 
in the broader context of quantum gravity, see e.g.~\cite{CW,FFZ}, 
in particularone is interested in the response of a QFT under an 
infinitesimal variation of the Unruh temperature. 
Of course the problem in the first place consists
in defining the ``continuation'' of the QFT in question. For the partition 
function Callan and Wilczek \cite{CW} proposed a ``replica'' 
prescription to define
its continuation. In \cite{MAX1} more generally the invariance of the 
S-matrix has been proposed as the defining criterion. Specifically 
for an integrable QFT this criterion turned out to uniquely determine the 
proper deformation of the form factor equations. In particular there is still
a cyclic equation whose cyclicity parameter $p \neq 2\pi i$ physically 
correponds to an off-critical Unruh temperature and which is equivalent 
to the qKZ equation at generic (non-zero) level. The main difference to
the ordinary form factor equations lies in the residue equations which 
prescribe how solutions with different number of variables are arranged 
into sequences. For models with a diagonal S-matrix the qKZ is easy to implement and 
some replica-deformed form factors have been computed in \cite{MAX1,MAP1}. 
However, the full set has not yet been found for any model. 

\medskip

As it was mentioned above the standard form factors of the 
Thirring model can be found among the solutions of (\ref{qKZ-def}) 
at level zero. It will be the key point of this paper to 
show that the physical problem of construction the 
``replica'' relatives of them can be solved by means 
of the solutions of (\ref{qKZ-def}) at generic level. 
To be a bit more precise we have to arrange solutions 
of (\ref{qKZ-def}) in sequences such that functions 
$\Psi(z_1, \ldots , z_n)$ with different numbers 
of arguments are linked by the form factor equations 
to be described in section 2. 

In particular if the local operator whose
form factors are being considered has isospin $j$ then in 
group theoreticalterms its $n$-particle form factor is an 
intertwiner $V_{1/2}^{\otimes n} \rightarrow V_j$,
where $V_j$ is the irreducible $sl_2$-module of isospin $j$.
In other words one has a one-to-one correspondence

\begin{equation}
\begin{tabular}{c} 
local operator of isospin $j$ \\  
in SU(2) Thirring model
\end{tabular}
\;\;\longleftrightarrow\;\; 
\begin{tabular}{c} 
sequence of level zero qKZ solutions\\ 
intertwining $V^{\otimes n} \rightarrow V_j,\;n \geq n_0$
\end{tabular}
\label{corr}
\end{equation}

Here $n_0$ is the starting member of a sequence and the non-vanishing 
members of a sequence have all either $n$ odd or $n$ even. 
For simplicity we shall refer to such a sequence as an isospin $j$ 
sequence. The aim in the bulk of the paper is to achieve something 
similar for the replica-deformed form factors. Since the 
replica-deformed system is no longer an ordinary Poincar\'{e} 
invariant QFT, the bootstrap viewpoint becomes even more important. 
That is the collection of deformed form factors is supposed 
to {\em define} the replica deformed system.

Our main result can then be summarized as follows: The right hand 
side of the correspondence (\ref{corr}) admits a unique 
replica-deformation, provided on both sides the equivalence classes 
modulo pointwise multiplication with a conserved charge eigenvalue 
(i.e.~a regular solution of the form factor equations 
with trivial S-matrix) are being considered. Symbolically we can 
write

\begin{equation}
\begin{tabular}{c} 
sequence of level zero \\
qKZ solutions of isospin $j$\\
modulo conserved charge eigenvalue
\end{tabular}
\;\;\longleftrightarrow\;\; 
\begin{tabular}{c}
sequence of level $p- 2\pi i \neq 0$ \\
qKZ solutions of isospin $j$\\ 
modulo conserved charge eigenvalue
\end{tabular}
\label{qcorr}
\end{equation}

The structure of the conserved charge eigenvalues in both cases is 
very different and so is the kinematical arena \cite{MAX1}. However,
in view of (\ref{qcorr}) the structure of the state space in both
theories is very similar as will be shown in section 5, 
much in the spirit of the replica idea \cite{CW,MAX1}.
   
\bigskip

The outline of the paper is as follows. In the next section 
we briefly review the replica-deformation of massive integrable 
field theories and describe the deformed form factor equations.
In section 3 we summarise the essentials of the bootstrap description 
of the Thirring model and prepare the deformed minimal 
form factors \cite{MAP1}. In section 4 we rewrite the solutions of 
the rational $sl_2$-type qKZ-equation of \cite{TV,MV} in a way 
that facilates the comparison with the (undeformed) form factors of 
the Thirring model. It turns out that the solutions can be 
parameterised by a polynomial similar to Smirnov's 
form factors. We proceed by generalising the completeness 
proof of the level zero solutions \cite{TAR} to generic level.
It is essentially this completeness result that underlies the 
correspondence (\ref{qcorr}). In section 6 we give the result for
the deformed form factors and first show that they satisfy a subset 
of the modified form factor equations of \cite{MAX1}. Here we take 
advantage of the fact that the replica deformation leaves the 
bootstrap S-matrix, and hence the structure of the algebraic 
algebraic Bethe vectors, unaffected. Then we derive conditions on the 
polynomial entering the qKZ-solutions under which the corresponding form 
factors in addition satisfy the modified kinematical residue equations. 
Finally, using (\ref{corr}) and (\ref{qcorr}) we give explicit 
expressions of some form factors which we conjecture to be 
the replica-deformed form factors of the Noether current in the 
SU(2) Thirring model. The last section is left to a discussion of the 
results.

\bigskip
\bigskip

{\section{Replica deformation of the form factor approach}

\bigskip
\bigskip

In this section we  review the results of \cite{MAX1,MAX2,MAX3} 
and state the modified form factor equations. 
The starting point is the Bisignano-Wichmann-Unruh
thermalisation phenomenon stating that 
the vacuum of  a Minkowski space QFT looks like a thermal state of 
inverse temperature $\beta = 2 \pi$ with respect to the 
Killing time of the Rindler wedge \cite{BW,UN}. This fact 
can be encoded symbolically in the following 
formula, where ${\cal{O}}_i(x_i)$, $i=1, \ldots, n$ denotes some local 
operator with support inside the Rindler wedge $W$ 

\beq
\langle 0 | {\cal{O}}_1(x_1) \cdots {\cal{O}}_n(x_n)  | 
0 \rangle = {\rm{tr}}\left[ \exp ( 2 \pi K ) 
 {\cal{O}}_1(x_1) \cdots {\cal{O}}_n(x_n) \right] .
\label{trace1}
\eeq

Here $K$ stands for the generator of Lorentz boosts in $W$. Note 
that this trace can -- in contrast to lattice models -- 
never exist in a continuum QFT due 
to the noncompactness of $K$ as described in \cite{MAX1,MAX3}. 
We therefore take (\ref{trace1}) only as a mnemonic.

Suppose now that we want to study ``the same'' QFT with the parameter
$\beta$ shifted away from $2\pi$, a problem which naturally
arises in the broader context of quantum gravity, see e.g.~\cite{CW,FFZ}.
Formally this amounts to replace

\beq
\exp ( 2 \pi K ) \rightarrow 
\left( \exp ( 2 \pi K ) \right)^{\beta/2\pi} = \exp( \beta K),
\label{repl}
\eeq
 
in (\ref{trace1}) while keeping everything else fixed. In the case 
of an integrable QFT we can require specifically that the factorized
S-matrix of the model is unchanged upon (\ref{repl}). This can be 
viewed as a concrete realisation of the ``replica'' idea of Callan and Wilczek 
\cite{CW}. In particular, if one could make sense out of 
the deformed correlator (\ref{trace1}), (\ref{repl}), its response
under an infinitesimal variation of the Unruh temperature could be
obtained simply by differentiation

\beq
\displaystyle{ 
\beta { { \partial}\over{\partial \beta}} 
 {\rm{tr}}\left[ \exp ( 2 \pi K ) 
 {\cal{O}}_1(x_1) \cdots {\cal{O}}_n(x_n) \right] \bigg|_{\beta = 2 \pi}.}
\label{trace2}
\eeq

The response (\ref{trace2}) again has a meaning within the context
of QFTs, while the deformed correlators themselves no longer 
correspond to a relativistic QFT. 

The crux of the problem of course is to define the deformed correlators
consistent with the replica idea (\ref{repl}) and in a way that avoids
mathematical ambiguities. For example naively regularizing the trace
in (\ref{trace1}) would most likely lead to enormous technical problems
with renormalizability since the translation invariance of the 
original QFT is lost for $\beta \neq 2\pi$. The solution proposed
in \cite{MAX1} is to implement the replica idea on the level of form 
factors.

\medskip

A form factor in this context is the matrix element 
of a local operator ${\cal{O}}(x)$ between the vacuum and an 
n-particle scattering state. The scattering states depend 
on rapidities $z_i$ and are created from the vacuum 
by means of Faddev-Zamolodchikov-type operators $A_{a_i} (z_i)$, 
where the index $a_i$ refers to the charge of a state. 
The form factor is then

\beq
F_{a_1\ldots a_n} (z_1, \ldots , z_n ) = 
\langle 0 | {\cal{O}}(x) | z_1, \ldots , z_n \rangle_{a_1\ldots a_n}
\sim
{\rm{tr}} \left[ e^{2 \pi K} {\cal{O}}(x) A_{a_1}(z_1) \cdots 
A_{a_n}(z_n) \right] ,
\label{ff-def}
\eeq

where we are again using our trace-mnemonic (\ref{trace1}).
In essence (\ref{ff-def}) says that a form factor always 
looks like a trace, a fact that has been derived from 
general quantum field theoretical principles and linked to 
the thermalisation phenomenon (\ref{trace1}) in \cite{MAX3}.  
Evidently the replica idea (\ref{repl}) can be applied to 
(\ref{ff-def}) as well, and here it can be made operational.
This is because for the replica deformed form factors 
a modified system of form factor equations exists.
It is uniquely determined by (\ref{repl}) and the requirement
that the S-matrix in unaffected by the deformation.
For fixed n the first two equations are as follows \cite{MAX1}. 

\beq
\beqcol
F_{a_1,a_2\ldots a_n} (z_1+ i \beta ,z_2, \ldots , z_n )  & = & \eta \;
F_{a_2\ldots a_n,a_1} (z_2, \ldots , z_n, z_1 ) ,    \\
F_{a_1,\ldots a_n, a_{n-1}} (z_1, \ldots , z_n, z_{n-1} )  & = & 
S_{a_{n-1} a_n }^{a_{n-1}^{\prime} a_n^{\prime}} (z_{n-1}-z_n )        
F_{a_1,\ldots,a_{n-1}^{\prime} a_n^{\prime} } (z_1, \ldots , z_{n-1}, z_n ).
\eeqcol
\label{watson}
\eeq
 
Here $S_{ab}^{cd}(z)$ is the factorised scattering operator. 
If this scattering operator has something to do with an 
R-matrix, it is not difficult to see that these two equations 
are equivalent to a qKZ equation (\ref{qKZ-def}) at generic level 
if we identify $p$ and $\beta$ in an obvious way. For $\beta =2\pi$ 
we recover 
the level zero situation or the standard Watson equations for form factors 
\cite{SMIR1}. The parameter $\eta$ is a phase, which may be different 
from $\eta =1$ for certain local operators of the model, 
see \cite{SMIR1,SMIR2}.

We need to supply the equations (\ref{watson}) with one more set of 
conditions. These conditions serve to single out those 
solutions of the qKZ equation which are of physical 
importance and provide a recursive relation between 
from factors involving different numbers of particles 
in the scattering state. We will refer to these additional 
conditions as kinematical residue equations.

We note that due to the first two requirements on the form 
factor (\ref{watson}) we may formulate without loss of 
generality the recursive relations just in the variables 
$z_{n-1}$ and $z_n$. 

The particular solutions of (\ref{watson}) searched 
for are according to \cite{MAX1} supposed to have 
simple poles at $z_{n}-z_{n-1}= \pm i \pi$ mod $i \beta$. 
The residues at these poles are determined by the 
follwing equations which link solutions of 
(\ref{watson}) with $n$ and $n-2$ variables.

\begin{eqnarray}
\res_{z_n = z_{n-1} + i \pi} 
F_{a_1,\ldots a_{n-1}, a_n} (z_1, \ldots , z_{n-1}, z_n )
&=& - C_{a_n a_{n-1}} 
F_{a_1,\ldots a_{n-2}} (z_1, \ldots , z_{n-2}).
\nonumber \\
\res_{z_n = z_{n-1} - i \pi} 
C^{a_{n-1} a_n} F_{a_1,\ldots a_{n-1}, a_n} (z_1, \ldots , z_{n-1}, z_n )
&=& - F_{a_1,\ldots a_{n-2}} (z_1, \ldots , z_{n-2}).
\label{KR1}
\end{eqnarray}

Here $C_{ab}$ is the charge conjugation matrix and $\lambda$ is a numerical
constant to be specified later. Note that the residue equation for 
$z_n = z_{n-1} + i \pi$ is similar to the condition satisfied by traces of 
vertex operators in lattice models \cite{JM1,JM2}. Without the second equation
(\ref{KR1}) however the recursion $n-2 \rightarrow n$ would be highly 
ambiguous. Taking advantage of the qKZ equation to implement the 
analytic continuation $z_n \rightarrow z_n + i\beta$ the second equation
(\ref{KR1}) can be rewritten as follows

\beq
\beqcol
\res_{z_n = z_{n-1} - i \pi + i \beta } 
F^{(n)}_{a_1\ldots a_n} (z_1, \ldots , z_{n-1}, z_n )
 & = &  \eta F^{(n-2)}_{b_1 \ldots b_{n-2}}(z_1, \ldots , z_{n-2}) 
\times \\
 & & {\hspace{-4cm}} \times C_{a_n c} 
 S_{c_1 a_1}^{c \;b_1}(z_{n-1} - z_{1} ) \cdots 
S_{a_{n-1}a_{n-2}}^{c_{n-1} b_{n-2}}(z_{n-1} - z_{n-2} ) .\\
\eeqcol
\label{KR2}
\eeq

This form turns out to be more convenient later. Observe that 
(\ref{KR2}) looks like the second term in the ordinary $\beta = 2\pi$
residue equation. For $\beta \rightarrow 2\pi$ the poles at 
$z_n = z_{n-1} + i\pi$ and $z_n = z_{n-1} - i\pi + i\beta$ merge and 
the residues add up, so that the ordinary residue equation of \cite{SMIR1}
is recovered.  

Equations (\ref{watson}) and (\ref{KR1}) are the replica deformed form factor 
equations. As explained in the introduction for a given S-matrix we take 
their solutions to define the replica deformed system. The variation of  
of $\beta$ away from $2 \pi$ then has wide ranging 
physical consequences. First of all the standard translation 
invariance is broken while Lorentz symmetry is maintained. 
Quite generally the kinematical properties of the $\beta \ne 2 \pi$ 
models are drastically changed. For example it has been 
shown in \cite{MAX1} that the mass eigenvalues of 
asymptotic states are altered in a way such that it costs more 
energy to boost two particles relative to each other 
than in the $\beta = 2 \pi$ case. Moreover, under certain 
circumstances not the entire momentum phase space is accessible 
to the particles. This phenomenon is close in spirit 
to 't~Hooft's picture of scattering states subject 
to quantum gravitational transmutation \cite{tH}.

\bigskip
\bigskip

{\section{The $SU(2)$-invariant Thirring model}}

\bigskip
\bigskip

The aim in the rest of the paper is to study the replica-deformation of the 
$1+1$-dimensional $SU(2)$-Thirring model in terms of its
form factors. In this section we prepare the basic ingredients for its bootstrap
description. For the sake of orientation however let us also display the 
classical Lagrangian. It is given in terms of a two-component 
spinor $\psi$, valued in the fundamental representation of $SU(2)$.

\beq
{\cal L} = i \bar{\psi} \gamma^{\mu} \partial_{\mu} 
\psi - g ( \bar{\psi} \sigma^{a} \gamma^{\mu} \psi ) 
 ( \bar{\psi} \sigma_{a} \gamma_{\mu} \psi )\,.
\label{lagrange}
\eeq

Here $\sigma^{a}$ are the Pauli matrices. The model exhibits dynamical mass 
generation and is believed to be asymptotically free.
The physical particles of the theory are an SU(2) doublet of 
massive kinks. The index $\varepsilon \in \{ \pm \}$ 
refers to either of these kinks. 

Moreover it is known that the model is integrable, and 
the factorised S-matrix is given as a function of the 
kink-rapidity $z$ by the following expression.

\beq
\displaystyle{
S_{\varepsilon_1 \varepsilon_2}
    ^{\varepsilon_1^{\prime} \varepsilon_2^{\prime}} (z) 
= S_0( z) { { z \;\; \delta_{\varepsilon_1}^{\varepsilon_1^{\prime}}     
                \delta_{\varepsilon_2}^{\varepsilon_2^{\prime}}
             + h \;\; \delta_{\varepsilon_1}^{\varepsilon_2^{\prime}}     
                \delta_{\varepsilon_2}^{\varepsilon_1^{\prime}} }
    \over
           { z+ h} } , \qquad h = - i \pi .}
\label{S-matrix}
\eeq

We use the abbreviation $h=-i \pi$ as indicated in (\ref{S-matrix}) 
from now on whenever suitable in accordance with \cite{NPT}. 
The scalar part $S_0$ of the S-matrix is given by

\beq
\displaystyle{
S_0 (z) = {  {\Gamma \left( { {1}\over{2}} - { {z}\over{2 h}} \right)
\Gamma \left( { {z}\over{2 h}} \right) } 
\over 
{\Gamma \left( { {1}\over{2}} + { {z}\over{2 h}} \right)
\Gamma \left(- { {z}\over{2 h}} \right) } }.}
\label{S0}
\eeq

Without $S_0$ the expression in (\ref{S-matrix}) coincides 
with the rational $sl_2$-R-matrix in the defining representation 
\cite{TV,MV}.

We also note that the charge conjugation matrix which enters 
the form factor equations (\ref{KR1}) and (\ref{KR2}) is 
given as follows.

\beq
C = i \sigma^2.
\label{charge}
\eeq

\medskip

The ordinary form factors associated with this S-matrix can be found
in \cite{SMIR1}. They have two features one expects to be present also
in the deformed case. First the tensor structure and second the appearence 
of a product of so-called minimal form factors in the solutions. 
Concerning the tensor structure we recall that an n-particle form factor 
(\ref{ff-def}) can be indexed by 
the kinks and anti-kinks of the scattering state and hence takes values
in an n-fold tensor product $V^{\otimes n} $ 
of fundamental representations of $SU(2)$. Using the Pauli matrices 
$\Sigma^{a} = \sum_{i=1}^{n}\sigma_{i}^{a}$ we can then 
define the weight subspaces 

\beq
( V^{\otimes n})_l = 
 \{ v \in  V^{\otimes n} | \Sigma^3 v = (n- 2 l) v \}.
\label{weight}
\eeq

We restict our attention 
to solutions $\psi (z_1, \ldots , z_n)$ of the qKZ equation valued in 
the space of singular vectors in $(V^{\otimes n})^{sing}_l $, 
i.e.~to those satisfying $\Sigma^{+} \psi (z_1, \ldots , z_n) = 0$. 
This is realted to an unbroken $SU(2)$-symmetry. Then $2 l \leq n$ and all
other components in an irreducible $SU(2)$-multiplet of isospin
$j = n/2 - l$ can be obtained by acting with $\Sigma^-$ on the 
singular vector. 
\medskip

To make the notation comparable to \cite{NPT,MV} we map the 
indices $a_1, \ldots , a_n$ of the form factor 
(\ref{ff-def}) to a set $M$ as follows. 
Let $M = \{ m_1 < m_2 < \ldots < m_l \} 
\subseteq \{ 1, 2, \ldots , n \}$, such that the 
number of elements of $M$ is $ \# M=l$. In other 
words this set labels the positions at which the 
form factor (\ref{ff-def}) has kinks of type ``$-$'' 
in the scattering state, i.e.~ $M= \{ i | \eps_i = - \}$. 
Hence, we can identify vectors in $V^{\otimes n}$ 
as 
\beq
v_M := v_{\varepsilon_1} \otimes \ldots \otimes 
        v_{\varepsilon_n}.
\label{vM-def}
\eeq
  
\medskip

Naturally we shall employ the weight decomposition (\ref{weight}) and 
the labelling (\ref{vM-def}) also in the deformed case. 

Further we prepare here the deformed counterpart of the minimal
form factor. This is (up to 
factors coming from representation theory and the 
pole structure) the two particle form factor as explained 
in \cite{MAP1}. The deformed minimal form factor $f(z)$ is 
a solution of the functional equations

\beq
f(z) = S_0 (z) f(-z), \qquad f(z+ i \beta) = f(-z).
\label{min-ff-1}
\eeq

The solution to these equations were found in \cite{MAP1} 
in terms of Barnes' Digamma functions. Let complex numbers 
$\omega_1, \omega_2$ be such that ${\rm{Re}}\; \omega_i >0 $ 
for $i=1,2$. We define the Digamma function \cite{JM2,MAP1} 
by the Hankel contour integral as 

\beq
\log \Gamma_2( x |\omega_1, \omega_2) =
\displaystyle{
{ {1}\over{2 \pi i}} \int 
{ { \exp ( - x t ) ( \log (-t) + \gamma) } \over
  { \prod_{i=1}^{2} ( 1 - \exp ( - \omega_i t) ) } }
{ { {\rm{d}}t} \over {t}}
},
\label{digamma}
\eeq

where $\gamma$ denotes the Euler constant. Note that the periods 
$\omega_1, \omega_2$ enter symmetrically in (\ref{digamma}). 
The Barnes' periodicity of this function is due to the 
relation 

\beq
\displaystyle{
{ {\Gamma_2( x+ \omega_1  |\omega_1, \omega_2)} \over 
  {\Gamma_2( x   |\omega_1, \omega_2)} } = 
{ {1}\over{\Gamma_1 (x|\omega_2)}},}
\label{period}
\eeq

where $\Gamma_1$ can be identified with the usual 
$\Gamma$ function as

\beq
\Gamma_1(x | \omega_2 ) = \omega_2^{-1/2 + x/\omega_2} 
\Gamma(x/ \omega_2) /\sqrt{2 \pi}.
\label{gamma-1}
\eeq

\medskip

The relevant solutions of (\ref{min-ff-1}) are then given by 

\beq
f(z) = 
\displaystyle{ 
{  { \Gamma_2 ( \pi - i z | 2 \pi, \beta ) 
     \Gamma_2 ( \pi + \beta + i z | 2 \pi, \beta )} \over 
  { \Gamma_2 ( - i z | 2 \pi, \beta ) 
     \Gamma_2 (  \beta + i z | 2 \pi, \beta )} }.
}
\label{min-ff}
\eeq

Note that apart from the higher periodicity 
the structure of (\ref{min-ff}) is similar to the one of the 
scalar S-matrix (\ref{S0}). This feature was observed to 
be common for a large class of integrable field theories 
in \cite{MAP1}. The minimal form factor (\ref{min-ff}) 
was given an interpretation in terms of central extensions 
of Yangian doubles in \cite{KLP}. As remarked before we can 
identify the ``inverse Unruh temperature'' $\beta$ with the shift 
parameter $p$ of the qKZ equation (\ref{qKZ-def}) via

\beq
p = -i \beta . 
\label{p-beta}
\eeq    

We find it convenient to use either of the variables throughout the paper, 
keeping the above identification in mind.
For $\beta= 2 \pi$ one recovers using (\ref{digamma}) from (\ref{min-ff}) 
up to trivial factors the minimal form factor 
$\zeta( z )$ used by Smirnov in \cite{SMIR1,SMIR2}. 

\medskip

For later use we conclude this section with the following 
identities among the minimal form factors

\beq
\beqcol
f(z) f(z- h) & = & \displaystyle{
 {1 \over{\Gamma_1( -i z | \beta )\Gamma_1( \beta - \pi +i z | \beta )}} } \\
     & &                                    \\
f(z) f(z+ h + i \beta ) & = & \displaystyle { { S_0 ( z)
 \over{\Gamma_1( i z | \beta )\Gamma_1( \beta - \pi -i z | \beta )}}  }\,.
\eeqcol
\label{minform-id}
\eeq
 
\bigskip
\bigskip

{\section{Solutions of the rational qKZ equation}}

\bigskip
\bigskip

It was mentioned in the previous section that -- up to 
the scalar factor -- the S-matrix of the Thirring model 
is just the rational $sl_2$-R-matrix in the defining 
representation. This means that the first two functional 
equations (\ref{watson}) to be satisfied by the form factor 
(\ref{ff-def}) can, -- up to scalar factors which turn out to 
be the minimal form factors -- be identified with the 
solutions of the qKZ equations with rational $R$-matrix 
as given in \cite{TV,MV}. Obviously our case is particularly 
simple in this respect because we have to work with only 
one type of representation of $sl_2$. 

Below we first recall the solutions of \cite{TV,MV} 
at arbitrary level and then proceed by rewriting them 
in a form suitable for our purposes. The new form 
readily allows one to make contact to the level zero solutions 
\cite{SMIR1,SMIR2,NPT} in the limit $p \rightarrow - 2\pi i$, 
and secondly allows for a transparent procedure to single 
out those qKZ-solutions which in addition satisfy the modified 
kinematical residue equations (\ref{KR1}) and (\ref{KR2}).

\medskip

We start by defining the rational hypergeometric space of \cite{TV} 
which will carry the information on the number and position 
of the kinks and anti-kinks of the form factor (\ref{ff-def}). 
This space consists of rational functions with at most 
simple poles and a certain asymptotic behaviour \cite{TV}. 
We define 

\beq
\hat{g}_M (t_1, \ldots , t_l ) = 
\displaystyle {
 \prod_{a=1}^{l} \left( { 1 \over { t_a - z_{m_a}- \Lambda} }
 \prod_{1 \le k < m_a } { { t_a - z_k + \Lambda} \over 
                          { t_a - z_k - \Lambda} }
                 \right).
              }
\label{ghat-def}
\eeq

Let $f = f(t_1, \ldots, t_l)$ be a function. Let $\sigma $ 
be an element of the symmetric group ${\bf S}^l$. 
Following \cite{TV,MV} we define a special 
symmetrisation in that for a simple transposition 
$(a,a+1) \in {\bf S}^l$ we set 

\beq 
[ f ]_{ (a,a+1)} (t_1, \ldots , t_a, t_{a+1}, \ldots, t_l) 
:= f(t_1, \ldots , t_{a+1}, t_a , \ldots , t_l) 
\displaystyle{ 
              {t_a-t_{a+1} + h} \over 
              {t_a-t_{a+1} - h}} .
\label{perm}
\eeq

Using this symmetrisation one can now introduce the following 
basis in the rational hypergeometric space. 

\beq
\hat{w}_M (t_1, \ldots , t_l) = 
\displaystyle{
          \sum_{\sigma \in {\bf S}^l} 
              [\hat{g}_M(t_1,\ldots , t_l)]_{\sigma}}.
\label{what-def}
\eeq
 
It is clear that in order to make our results comparable 
with \cite{SMIR1,SMIR2} we have to get rid of the unusual 
symmetrisation occuring in (\ref{what-def}).

An important object in the theory qKZ equations is the phase function. 
This object provides the link between the analytic structure 
of the solutions and the geometry of local systems \cite{TV,EFK}. 
In the conventions of \cite{TV} the phase function is given by:

\beq
\hat{\Phi}( t_1, \ldots, t_l; z_1, \ldots , z_n) = 
\displaystyle{ 
        \prod_{a=1}^{l} \prod_{j=1}^{n} \hat{\phi}( t_a - z_j; \Lambda) 
  \;
        \prod_{1 \le a < b \le l} \hat{\phi}( t_a - t_ b; h ), 
}
\label{phase-def-TV-1}
\eeq

where 

\beq
\displaystyle{ 
\hat{\phi}( x ; \alpha) := { {\Gamma( (x+ \alpha)/p) } \over 
                             {\Gamma( (x- \alpha)/p) } }
}.
\label{phase-def-TV-2}
\eeq
   
We refrain from describing the connection coefficients arising 
from (\ref{phase-def-TV-1}), details can be found in \cite{TV}.

The third and last entity to describe the solution spaces of 
the rational qKZ-equation is the trigonometric hypergeometric 
space ${\cal F}_q$. The functions in this space are of the form

\beq
P(\xi_1, \ldots , \xi_l ; \zeta_1, \ldots , \zeta_n) 
\displaystyle{
\prod_{a=1}^l \prod_{j=1}^n { { {\rm{exp}} (i \pi (z_j - t_a) /p ) }
             \over { \sin ( \pi ( t_a-z_j -\Lambda)/p) }}
\prod_{1\le a < b \le l} { {\sin( \pi ( t_a - t_b)/p) } \over 
                         {\sin( \pi ( t_a - t_b - h)/p) }}.
}
\label{W-def}
\eeq

Here we have set $\xi_a = {\rm{exp}}( 2 \pi i t_a/p)$ and 
$\zeta_j = {\rm{exp}}( 2 \pi i z_j/p)$. $P$ is a polynomial. 
It will be shown in the next section that the completeness 
of the solutions is encoded into the properties of $P$. 
Moreover, the local operators in the model can be 
classified in terms of this object.

\medskip

We can now write down the space of solutions of the qKZ 
equation for a function with values in $V^{\otimes n}$ in terms 
of an integral representation. Let 
$W= W_P$ be a function of the space ${\cal F}_q$ defined 
in (\ref{W-def}). The integrand is given by

\beq
I(\hat{w}_M, W_P ) = \hat{w}_M (t_1, \ldots , t_l) 
\hat{\Phi} (t_1, \ldots , t_l) 
W_P (t_1, \ldots , t_l), 
\label{integrand}
\eeq

where we omitted the dependence of $I(\hat{w}_M, W_P )$ on 
$z_1, \ldots , z_n$. The solution of the equation (\ref{qKZ-def}) 
with values in representations indicated in the previous 
section is then given by:

\beq
\Psi_{W}(z_1, \ldots , z_n)  =  
\displaystyle{ 
\sum_{\# M = l} \int_{C_1} \cdots \int_{C_l} 
{\rm d}t_1 \cdots {\rm d}t_l \; I ( \hat{w}_M, W_P )} \; v_M .
\label{TV-sol}
\eeq

We specify the integration cycles $C_l$ 
below after having rewritten this solution in a more convenient 
form. It has been shown in \cite{TV,MV} that $\Psi_W$ is a 
solution to the qKZ equation. We will indicate how to 
check this after having introduced a basis in the 
solution space which is more appropriate to the aim 
of this paper. We stress again that the information on the space of 
representations 
of $sl_2$ is contained only in the function $\hat{w}_M$. 
Physically speaking this means that once we have fixed the 
number of kinks and anti-kinks in the candidate form factor, the 
only freedom of choice resides in the polynomial $P$. 

\medskip

Eventually we are interested in constructing the 
form factors of the replica-deformed $SU(2)$-invariant Thirring 
model. To this end it is useful to rewrite the 
solution spaces (\ref{TV-sol}) in a form which allows 
a direct comparison with the standard level zero form factors 
as they have been given in \cite{SMIR1,SMIR2,NPT}. In 
particular we need to get rid of the non-standard 
symmetrising operation (\ref{perm}) for the rational 
weight function (\ref{what-def}). However, the alterations 
to be performed do not change (up to an overall normalisation) 
the solution (\ref{TV-sol}). 

We replace some objects appearing in (\ref{TV-sol}) as follows: 

\beq
t_a - z_j \rightarrow t_a - z_j + h, \quad \forall a, j;
\qquad \Lambda \rightarrow -h/2 .
\label{replace}
\eeq

This replacement will be in effect from now on. 

Next, we replace a part of the phase function (\ref{phase-def-TV-1}) 
which does depend only on differences of the variables $t$ as follows.

\beq
\displaystyle{
{ {1}\over { \Gamma \left( { { t_a-t_b -h} \over {p}} \right)}} 
\sim 
 \Gamma \left( { { h-(t_a-t_b)} \over {p}} \right)
( t_a - t_b -h) \sin{ {\pi}\over{p}}(t_a - t_b -h) }.
\label{schritt1}
\eeq

As it is implicit already in \cite{NPT} we can use this 
to replace the non-standard symmetrisation (\ref{perm}) 
by a standard one.
\beq
w_M (t_1, \ldots , t_l ) : = \hat{w}_M (t_1, \ldots , t_l) 
        \prod_{1 \le a < b \le l} ( t_a - t_b -h) 
 = {\rm Asym}\left( g_M (t_1, \ldots , t_l) \right), 
\label{w-def}
\eeq

where 

\beq
g_M (t_1, \ldots , t_l) = \hat{ g}_M (t_1, \ldots , t_l) 
          \prod_{1 \le a < b \le l} ( t_a - t_b -h),
\label{g-def}
\eeq

and Asym denotes the standard antisymmetrisation with 
respect to the variables $t_1, \ldots , t_l$. 

This means that upon performing the steps outlined above, 
we can work with the same rational weight function as in 
the level zero case \cite{NPT}. The same feature has 
been used in \cite{MT} for determinant formulas in the 
case of the trigonometric qKZ equation.  

As will be explained below, the natural consequence of this is 
that the (algebraic) Bethe vectors in the $\beta$-deformed case 
will be the same as in the standard level zero case. 

In addition we can also cancel the pure 
$t$-dependent part in the denominator of (\ref{W-def}). 
This has in particular the remarkable consequence that the space 
${\cal F}_q$ introduced in (\ref{W-def}) almost coincides with 
the corresponding space introduced in \cite{NPT,TAR} in 
the level zero case. We shall exploit this fact below.

\medskip

Next we reorganise the remaining parts of 
the integrand in (\ref{TV-sol}). This will be done in a 
way to point out the similarity between (\ref{TV-sol}) 
and trace formulas of vertex operators as developed 
in \cite{JM1,JM2,KLP}. To this end we first define the analogue 
of the phase function used 
by Smirnov \cite{SMIR1,SMIR2}.

\beq
\Phi(y ) := 
\displaystyle{ 
\Gamma \left( { {  h/2 + y  } \over { p}} \right ) 
\Gamma  \left( { { p- 3 h/2- y } \over { p}} \right ) 
\exp ( - { { \pi i} \over {p}} y ).
}
\label{phi-def}
\eeq

Note that here we put in the exponential factor from 
(\ref{W-def}) just as a matter of convenience.

Then we define an odd function necessary to guarantee the 
complete symmetry of the integrand with respect to $t$. 

\beq
\psi (y ) := \displaystyle{
 { {-1}\over {  \pi^3 p}}
 \Gamma \left( { { h + y }\over {p}} \right)
 \Gamma \left( { { h - y }\over {p}} \right)
  \sin ( { {\pi}\over {p} }\;  y ).
}
\label{psi-def}
\eeq

For later use let us note the identity

\beq
\Phi(y) \Phi(y+h) = { {  \; \exp (-2\pi i(y+ h/2)/p) } \over
{ \psi(y+ 3 h/2) ( y+ h/2) \sin{ {\pi}\over{p}}(y+ h/2 )
\sin{ {\pi}\over{p}} (y+ 5 h/2 )  }}.
\label{phi-psi-id}
\eeq

Observe  that this expression (up the the exponential factor) 
tends to the corresponding identity for the phase functions 
in \cite{SMIR1,SMIR2} in the case $\beta \to 2 \pi$.

Having now collected all the data we can write down the solution 
of the qKZ equation in a new form, which is, however, up to 
a normalisation identical with (\ref{TV-sol}). Since now 
the only degrees of freedom of the solution reside in the 
polynomial $P$ of (\ref{W-def}) we prefer to label the 
solution by this object.

\beq
\beqcol
\Psi_P(z_1, \ldots , z_n) & = &  
\displaystyle{ \sum_{\#M = l}
\int_{C_1} \cdots \int_{C_l}  
w_M(t_1, \ldots , t_l) 
P (\xi_1, \ldots, \xi_l; \zeta_1, \ldots , \zeta_n)\; \times 
} \\
& \times & \displaystyle{ 
\prod_{a=1}^{l} \prod_{j=1}^{n} \Phi( t_a - z_j ) 
\prod_{1 \le a < b \le l} \psi (t_a - t_b) \;
{\rm d}t_1 \cdots {\rm d}t_l \; v_M
}.\\
\eeqcol
\label{qKZ-sol}
\eeq

The singular hyperplanes of the integrand in (\ref{qKZ-sol}) 
are 

\beq
\beqcol
t_a & = & z_j - h/2 + {\bf Z}_{\le 0} p, \\
t_a & = & z_j + p - 3 h/2 + {\bf Z}_{\ge 0} p, \\
t_a & = & z_j - 3 h /2, \quad {\rm depending \quad on} \quad M, \\
t_a & = & t_b -h + {\bf Z}_{\le 0} p , \\
t_a & = & t_b + h + {\bf Z}_{\ge 0} p .
\eeqcol
\label{poles}
\eeq

Note that the comment in the third line means that this pole 
may occur only in the basis function $w_M$ and hence its presence 
depends on the indexing set $M$.

This now allows us to specify the integration contours as follows. 
For any $a=1, \ldots , l$ we let the integration contour 
$C_a$ separate  

{\it i.} the hyperplanes of the first line from 
the ones of the second and third line in (\ref{poles}).

{\it ii.} the hyperplanes of the fourth line from the 
ones in the fifth line of (\ref{poles}).

\medskip

This choice is convenient for our purposes, but as it has been 
indicated in \cite{TV,MV} the solution space is by analytical 
continuation actually independent of the particular choice 
of the integration contour.

We are now going to check that (\ref{qKZ-sol}) is a solution 
of the rational qKZ-equation. Of course this is clear 
by construction \cite{TV}. However, along the way we introduce 
some new objects which will be of importance later in the 
paper. The use of these objects is nicely illustrated 
by verifying  the qKZ equations. 

A fundamental role in \cite{TV} is played by discrete 
derivates with respect to the variables $t$ and $z$ 
respectively. We describe these derivatives within the 
modified form of the solution space (\ref{qKZ-sol}). 
Therefore we introduce shift operators:

\beq
\beqcol
Z_j f(t_1,\ldots , t_l; z_1, \ldots, z_n) & := & 
\phi_{l+ j}({\bf t}, {\bf z}) \;
f(t_1, \ldots , t_l; z_1, \ldots , z_j+p ,\ldots , z_n), \\
& &                                                       \\
Q_a f(t_1,\ldots , t_l; z_1, \ldots, z_n) & := & 
\phi_{a}({\bf t},{\bf z}) 
f(t_1, \ldots, t_a + p ,\ldots,  t_l; z_1,\ldots , z_n), 
\eeqcol
\label{shift-def}
\eeq

where 

\beq
\beqcol
\phi_{l+ j}({\bf t}, {\bf z}) & = &
\displaystyle{ \prod_{a=1}^{l}  { { t_a - z_j + 3 h/2 -p} \over
                             { t_a - z_j +  h/2 -p}} } , \\
& &                                    \\
\phi_{a}({\bf t}, {\bf z}) & = & \displaystyle{ 
\prod_{{b=1} \atop{b \ne a}}^{l} { { t_a - t_b + h} \over 
                                  { t_a - t_b + p -h}} \;
\prod_{j=1}^{n} {  {t_a-z_j+ h/2} \over {t_a -z_j + 3 h/2} }} ,
\eeqcol
\label{phasefac}
\eeq

are the connection coefficients of \cite{TV,MV} adjusted to 
the choice of the solution (\ref{qKZ-sol}).

With these objects we can define discrete partial derivatives, 
using the convention that $D_a$ for $a= 1 , \ldots , l$ acts 
on variable $t_a$ while $D_{l+j}$ for $j=1, \ldots , n$ acts 
on variable $z_j$. 

\beq
D_a = {\bf 1} - Q_a, \qquad D_{l+j} = {\bf 1} - Z_j.
\label{der-def}
\eeq

The proof that (\ref{qKZ-sol}) is a solution of the rational 
qKZ equation is then parallel to the one given in \cite{NPT} 
for the level zero case. We state the necessary 
steps here because we will need this result later for 
the solution of the first two form factor equations 
(\ref{watson}). 

Following (\ref{S-matrix}) and (\ref{S0}) let 
$R(z)^{\varepsilon^{\prime}_1 \varepsilon^{\prime}_2}_{
           \varepsilon_1 \varepsilon_2}= 
S(z)^{\varepsilon^{\prime}_1 \varepsilon^{\prime}_2}_{
           \varepsilon_1 \varepsilon_2} S_0(z)^{-1} $, and 
$f^{(a)}_{\varepsilon_1, \ldots , \varepsilon_n}$ for 
$a=1, \ldots , l$ be a family of functions which can be 
expressed in terms of the rational hypergeometric basis 
functions (\ref{w-def}). The qKZ equation for functions 
of the form (\ref{qKZ-sol}) valued in $V^{\otimes n}$ is 
then equivalent to the following equations on the 
rational hypergeometric basis functions (\ref{w-def}), taking 
into account the definition of the set $M$ at the end of 
section 3. 

We state the first one in $z_1$ only, but its generality is 
obvious. 

\beq
Z_1 w_{\varepsilon_1, \varepsilon_2, \ldots , 
\varepsilon_n} ( \cdot; z_1, z_2, \ldots , z_n )
-  w_{\varepsilon_2, \ldots, \varepsilon_n, 
\varepsilon_1} (\cdot; z_2,  \ldots , z_n ,z_1) 
= \sum_{a=1}^l D_a f^{(a)}_{\varepsilon_1, \ldots , \varepsilon_n}.
\label{qKZ-proof-1}
\eeq

The second equation refers to the exchange of two $z$ variables

\beq
\beqcol
w_{\varepsilon_1,\ldots , \varepsilon_{i+1}, \varepsilon_i, \ldots 
\varepsilon_n} (\cdot ;z_1, \ldots, z_{i+1}, z_i, \ldots , z_n ) 
& = &                                                     \\ 
& & {\hspace{-6cm}}
\displaystyle{
\sum_{\varepsilon^{\prime}_1,\varepsilon^{\prime}_2= \pm}
 R^{\varepsilon^{\prime}_i \varepsilon^{\prime}_{i+1}}_{
           \varepsilon_i \varepsilon_{i+1}}(z_i-z_{i+1}) 
w_{\varepsilon_1,\ldots , \varepsilon^{\prime}_{i}
\varepsilon^{\prime}_{i+1}, \ldots , 
\varepsilon_n} (\cdot ;z_1, \ldots, z_{i}, z_{i+1}, \ldots , z_n ) .}
\eeqcol
\label{qKZ-proof-2}
\eeq

We then have the following

\bigskip

{\bf Theorem 1}: {\it The function (\ref{qKZ-sol}) valued 
in the tensor product $V^{\otimes n}$ is a solution 
of the rational $sl_2$ qKZ-equation.}

\medskip

{\sc Proof:} For the proof it is sufficient to recall that 
the hypergeometric integral of \cite{TV,MV} as defined in 
(\ref{qKZ-sol}) has exact hypergeometric forms ({\it cf.}~\cite{TV} 
Lemma 2.21), which are essentially given 
by the derivatives $D_a$ defined in (\ref{der-def}).

\bigskip
\bigskip

{\section{Completeness of the solutions}

\bigskip
\bigskip

Here we prove the completeness of the solutions (\ref{qKZ-sol}) 
of the qKZ equation. On the basis of form factors in the Sine-Gordon 
model the completeness of the solutions was shown by Smirnov 
\cite{SMIR3} in establishing a bilinear relation among the 
solutions of the form factor space. Later is was shown by 
Tarasov \cite{TAR} that (at least for the rational qKZ 
equation) similar relations can be extracted from the 
solution spaces of the qKZ equation directly. In \cite{TAR} 
the fact was used that for the form of the qKZ equation 
(\ref{qKZ-def}) studied in this paper, there exist 
zero solutions. In our language this means that there 
exist polynomials $P$ in (\ref{qKZ-sol}) such that the 
integral vanishes automatically. This guarantees that the 
dimension of the solution space matches the dimension 
of $(V^{\otimes n})_{sing}$ rather than that of $V^{\otimes n}$. 
For a detailed presentation of this topic, see \cite{TV}. 

In this section we are going to derive a completeness 
relation for the solutions of the qKZ-equation at generic 
level, similar to the one in the level zero case \cite{SMIR3,TAR}.
Since we modelled our solution parallel 
to the level zero case \cite{NPT,TAR} we can expect that the 
proofs here will be quite similar to the ones given in 
\cite{NPT,TAR}. The main difference is due to the more 
complicated pole structure of the integrand (\ref{poles}).

We begin by noting the dimension of the space ${\cal F}_q$ introduced in 
(\ref{W-def}); it is finite dimensional with ${\rm dim}{\cal F}_q = 
{\rm dim}(V^{\otimes n})_l = {n \choose l}$. 

Let us first consider the case $l=1$ in the integral (\ref{qKZ-sol}), 
which means in particular that there are no $\psi$-functions 
present. Insert for the polynomial $P$ either of the two choices 

\beq
\displaystyle{
P_{s1}(t_1) = e^{-2\pi i h n/p} \prod_{j=1}^n \exp{ {\pi}\over{p}} ( t_1 - z_j + 3 h/2) , 
\quad {\rm{or}} \quad P_{s2}(t_1) =
\prod_{j=1}^n \exp{ {\pi}\over{p}} ( t_1 - z_j + h/2) .}
\label{ker-1}
\eeq

If we ignore for the moment that the partial degree in 
$\exp(2\pi i t_1/p)$ is not as it may have been assumed in (\ref{W-def}) 
it is not difficult to check that the first expression 
cancels all possible poles appearing in the second line of 
(\ref{poles}). Closing the contour $C_1$ in (\ref{qKZ-sol}) 
such that it encircles the poles in question, we realise 
that the integral vanishes. 

The same kind of argument applies to the second choice 
in (\ref{ker-1}). Here we close the contour $C_1$ such that 
it contains all possibles poles appearing in the first 
line of (\ref{poles}). 

Hence we have established that for $l=1$

\beq
\Psi_{P_{s1}} = 0 =\Psi_{P_{s2}}.
\label{kernel1}
\eeq

This result is the generalisation of Lemma 5.3 in \cite{NPT} to the 
generic level case.

\bigskip

Let us proceed to the case of $l=2$. According to \cite{TAR} 
we set 

\beq
\Xi^{(1)}(t) = P_{s_1}(t) - P_{s_2} (t).
\label{xi1-def}
\eeq

From the partial degree point of view this is a proper candidate for 
the polynomial $P$. It is clear from (\ref{kernel1}) that we also have 

\beq
\Psi_{\Xi^{(1)}} =0.
\label{kernel1-1}
\eeq

Let us give an interpretation of this object. According to 
\cite{TV} it is possible to construct basis functions 
similar to (\ref{w-def}) for the trigonometric hypergeometric 
space as defined on (\ref{W-def}). If we take the definition 
(2.26) of \cite{TV} and rewrite the basis according to our 
form of the solutions (\ref{qKZ-sol}) of the qKZ-equation, we 
find that these basis functions for $l=1$ and 
variables $z_1, \ldots , z_n$ can be recast 
as a polynomial

\beq
P_{\mu}(t) =\displaystyle{
 \exp\left(4\pi i \Lambda (n -\mu)/p \right)
 \prod_{1 \le j < \mu} \left( e^{2 \pi i ( t- z_j + \Lambda)/p} -1 
           \right)
\prod_{\mu <  j \le n} \left( e^{2 \pi i ( t- z_j - \Lambda)/p} -1 
           \right).}
\label{P-1}
\eeq

Note that we still keep the convention (\ref{replace}). 
We can formally extend the definition (\ref{P-1}) also 
to the case $\mu=0$. 

Then it is straightforward to check that

\beq
\displaystyle{
\prod_{j=1}^{n} \left(e^{2 \pi i ( t- z_j + \Lambda)/p} -1 \right) 
= P_0 (t) + (e^{4\pi i \Lambda/p} -1) \sum_{j=1}^{n}P_j(t).}
\label{P-2}
\eeq
 
The trigonometric hypergeometric basis functions for $l=2$ 
in their polynomial form (appropriate to the conventions 
laid out in the previous section) can for $\mu < \nu$ be 
constructed from (\ref{P-1}) as 

\beq
P_{\mu\nu}(t_1,t_2) = 
\left( \sin{ {\pi}\over{p}} (t_1-t_2 ) \right)^{-1} 
{\rm{Asym}}\;\left( P_{\mu} (t_1) P_{\nu} (t_2) 
\sin{ {\pi}\over{p}}(t_1-t_2 -h) \right).
\eeq

Having seen the structure of the basis functions, we can now 
write down a polynomial $\Xi^{(2)}$ 

\beq
\Xi^{(2)} (t_1, t_2) = \Xi^{(1)}(t_1) \Xi^{(1)}(t_2) 
\left( \sin{ {\pi}\over{p}} (t_1-t_2 ) \right)^{-1} 
\left( \sin{ {\pi}\over{p}}(t_1-t_2 -h) - 
       \sin{ {\pi}\over{p}}(t_2-t_1 -h)\right).
\label{xi2-def}
\eeq

Taking into account the pole structure (\ref{poles}) it is 
possible along the lines in the $l=1$ case to check that 
for $l=2$ we have

\beq
\Psi_{ \Xi^{(2)} } = 0.
\label{kernel2}
\eeq

The functions $\Xi^{(1)}$ and $\Xi^{(2)}$ are the analogs 
of the kernel functions at level zero as introduced by 
Tarasov in \cite{TAR}. Therefore the construction of the 
graded spaces in section 7 of \cite{TAR} can be applied 
to the case of generic level as well.

This means that we have now found the necessary set of kernel 
polynomials for the solutions of the qKZ-equation. The 
function $\Xi^{(2)}$ is the analog of what was found in 
\cite{SMIR3} as a consequence of the deformed Riemann 
bilinear identity at level zero. 

We say 
that the arguments $z_1, \ldots , z_n$ are in generic position, 
if the singular hyperlanes in (\ref{poles}) do not coincide. 
For the case when they do, see section 6.

The main point now is that the kernel solutions (\ref{kernel1-1}) 
and (\ref{kernel2}) reduce the actual dimension of the parameter 
space from dim$(V^{\otimes n})_l$ to 
dim$(V^{\otimes n})^{sing}_l$. This can be established 
using arguments absolutely parallel to the ones employed in 
the proof of Theorem 4.3 and 4.4 in \cite{TAR}. This then gives 
rise to the following

\bigskip

{\bf Theorem 2:} {\it For generic values of the arguments 
$z_1, \ldots , z_n$, the solutions (\ref{qKZ-sol}) span 
the space $(V^{\otimes n})^{sing}_l$. }

\bigskip

{\it Remark}: This result is crucial for the correspondence 
(\ref{qcorr}) between the ordinary and the deformed form factors.
We shall return to it after Theorem 4. But let us indicate 
here that it means that the number of physical states in the 
replica-deformed model is identical to the one in the 
standard case. A fact which, however, could have been 
anticipated.

\bigskip
\bigskip

{\section{The replica-deformed form factors}

\bigskip
\bigskip

In this section we define the replica-deformed 
form factors and check under which conditions on the 
objects $P$ they satisfy the modified form factor 
equations of section 2. According to what has been said in section 3 
we can encode the index structure of the form factor (\ref{ff-def}) as 

\beq
F_{\eps_1, \ldots , \eps_n}(z_1, \ldots , z_n ) = 
F_M(z_1, \ldots , z_n ).
\label{eps-M}
\eeq

We denote by ${\cal{I}}^P_M(z_1, \ldots , z_n)$ the summand 
corresponding to the basis element $w_M$ in the expression 
for the solution $\Psi_P$ of the qKZ-equation, as stated in 
(\ref{qKZ-sol}).

Recall that the form factors of the standard Thirring model 
\cite{SMIR1,SMIR2} are ({\it cf.} \cite{NPT}) spe\-cial solutions 
of the rational $sl_2$ qKZ equation at level zero multiplied 
by a scalar function. We will now show that the form factors 
in the replica-deformed Thirring model in the sense of \cite{MAX1} 
naturally arise from the solutions of the rational $sl_2$ qKZ 
equation at generic level. Specifically we will show
that for a suitable choice the objects $P$ the functions

\beq
\displaystyle{
F_M(z_1, \ldots , z_n) = { {c_l}\over { (2\pi i)^l}} 
\prod_{n \ge i > j \ge 1} f(z_i - z_j) \times
{\cal{I}}^P_M(z_1, \ldots , z_n),}
\label{sol-1}
\eeq

are solutions of the deformed form factor equations 
(\ref{watson})--(\ref{KR2}). Note that we take 
$2 l \le n$, later we shall see that $j=n/2 -l$ can be identified 
with the isospin of the underlying local operator. 
For the constant $c_l$ we take:

\beq
c_l = \left( 
2 \pi h \; \Phi( - 3 h/2) \Gamma (-h/p) (\beta)^{\pi/ \beta} 
e^{i \pi h / (2 p) } \right)^{-l}.
\label{cl-def}
\eeq

\medskip

The fact that (\ref{sol-1}) is a solution to the first two 
form factor equations (\ref{watson}) can be proved without 
problems. Due to its importance we state this fact as the 
following

\bigskip

{\bf Theorem 3}: {\it The function $F_M(z_1, \ldots , z_n)$ 
as defined in (\ref{sol-1}) satisfies the form factor equations 
(\ref{watson})}.

\bigskip

{\sc Proof:} Take the identities (\ref{min-ff-1}) for the 
minimal form factor $f(z)$ to realize that the factor 
$\prod_{n \ge i > j \ge 1} f(z_i - z_j)$ saturates the 
scalar part of the $S$-matrix (\ref{S-matrix}) in the 
form factor equations (\ref{watson}). The Theorem in 
section 4 together with (\ref{qKZ-proof-1}) and 
(\ref{qKZ-proof-2}) establishes the result.

\bigskip
\bigskip

{\it Remark:} If we stick to the definition of the form factor 
given in (\ref{sol-1}), which means that we keep the the 
requirements set up earlier on the polynomial $P$ in 
(\ref{W-def}), then $F_M$ satisfies (\ref{watson}) with 
$\eta=1$. There are, however, situations where we have 
to weaken the requirements on $P$ depending on the 
local operator we are going to describe with this object. 
For example in the next section we will see an example, 
where we relax the $p$-periodicity in the variables 
$z$ of $P$. In this case $\eta$ may be different from 
unity but still satisfies $ | \eta | =1$.

\bigskip

We now come to the part in the form factor 
business going beond the qKZ-equation. Namely, we 
have to single out those solutions described in Theorem 
3, which do satisfy the equations (\ref{KR1}) or (\ref{KR2}). 

Our strategy to prove this will be similar to the one used 
by Smirnov \cite{SMIR1,SMIR2}. In our case this means the 
following. In \cite{SMIR1,SMIR2} the source for the poles 
$z_i=z_j-h$ were taken to reside entirely in the algebraic 
Bethe vectors $\omega_M$. Additional contributions arising from the 
residue evaluation of the integrals (\ref{qKZ-sol}) at level 
zero, the origin 
of which is the pinching of integration contours, are removed by a 
trick. 

The algebraic Bethe vectors $\omega_M$ were shown in \cite{NPT} 
(Lemma 6.4) to descend directly from the rational basis functions 
$w_M$. Now we have seen that at generic level by (\ref{w-def}) 
we can work with the same basis functions as in the level zero 
case. 

This essentially reflects the {\it replica}-understanding of \cite{CW} 
in varying the Unruh temperature in an integrable qauntum field 
theory as sketched in section 2. There we have seen that this 
thermalisation changes many aspect of the physics but preserves 
the $S$-matrix of the zero temperature model. And as the $S$-matrix 
is unchanged, the algebraic Bethe ansatz \cite{SMIR2} remains, 
of course, unchanged. We could now outline along the lines 
of \cite{NPT} how to construct the Bethe vectors from the 
the solution of the qKZ-equation (\ref{qKZ-sol}). However, 
this is quite technical and in addition we would like 
to show that the kinematical residue equations 
(\ref{KR1}) and (\ref{KR2}) can be solved using the 
basis functions $w_M$ directly and using 
the singular hyperpalnes of the hypergeometric integral 
(\ref{poles}) only (mainly because the other 
way is by Smirnov's work very well understood). 

For that purpose we simply state 
the now obvious but physically important 

\bigskip

{\bf Lemma 1:}  {\it The replica deformation of the $SU(2)$-invariant 
Thirring model has no effect on the algebraic Bethe ansatz.}
   
\bigskip
\medskip

We can now proceed to find solutions of the kinematical 
equations (\ref{KR1}) and (\ref{KR2}). We first collect some 
results on the residues of the functions 
$w_M(t_1, \ldots , t_l; z_1, \ldots , z_n)$ as defined 
in (\ref{w-def}) as they will be needed to evaluate 
some of the integrals ${\cal{I}}_M^P$ below. 
For the proofs to be given below it 
is sufficient to consider only poles in the variables 
$z_{n-1}$ and $z_n$. Therefore we prefer not to write 
down most of the formulas in the sequel in full generality, 
but rather stick to the sufficient case.

For both $n-1, n \nin M$ or $n-1, n \in M$ we find that
\beq
\res_{t_l= z_n- 3 h/2 }w_M(t_1,\ldots, t_l) = 0 = 
\res_{t_l= z_{n-1}-3 h/2}w_M(t_1, \ldots , t_l) .
\label{res-form-1}
\eeq

The next case is when $ n \in M$ but $n-1 \nin M$. We set 
$N= M \setminus \{n \} $.

\beq
\beqcol
\displaystyle{
\res_{t_l = z_n-3 h/2} w_M(t_1,...,t_l; z_1, ... , z_n) |_{z_n=z_{n-1}-h} }
 & = & 2 \; w_N( t_1, ... , t_{l-1}; z_1, ... , z_{n-2} ) \times  \\

& & {\hspace{-3cm}}  
\times \displaystyle{\prod_{k=1}^{n-2} { {z_{n-1}- z_k - 2 h}\over 
                          {z_{n-1}- z_k -  h} } 
\prod_{a=1}^{l-1} (t_a - z_{n-1} + 3 h/2 ) },          \\
\displaystyle{
\res_{t_l = z_n- 3 h/2}} w_M(t_1,...,t_l; z_1, ... , z_n) |_{z_n=z_{n-1} +h} 
& = & 0 . 
\eeqcol
\label{res-form-2}
\eeq

If we take the residue with respect to $z_{n-1}$ rather than $z_n$, we get:

\beq
\beqcol
\displaystyle{\res_{t_l = z_{n-1}-3 h/2}
 w_M(t_1,...,t_l; z_1, ... , z_n) |_{z_n=z_{n-1}-h}} 
 & = & - w_N( t_1, ... , t_{l-1}; z_1, ... , z_{n-2} ) \times \\
& &  {\hspace{-3cm}}  \times 
\displaystyle{ \prod_{k=1}^{n-2} { {z_{n-1}- z_k - h}\over 
                          {z_{n-1}- z_k} } 
\prod_{a=1}^{l-1} (t_a - z_{n-1}+h/2 ) },          \\
\displaystyle{\res_{t_l = z_{n-1}-3h/2} } 
w_M(t_1,...,t_l; z_1, ... , z_n) |_{z_n=z_{n-1} +h} 
& = &               \\             \\
 & &  {\hspace{-2cm}} = - \displaystyle{\res_{t_l = z_{n-1}-3 h/2}} 
w_M(t_1,...,t_l; z_1, ... , z_n) |_{z_n=z_{n-1}-h}  . 
\eeqcol
\label{res-form-3}
\eeq

At last we treat the case of $n \nin M$ but $n-1 \in M$. Set $N^{\prime} 
= M \setminus \{ n-1 \}$. Here we 
find that $\res_{t_l = z_n-3 h/2 } w_M(t_1,\ldots , t_l) = 0$, while

\beq
\beqcol
\displaystyle{
\res_{t_l=z_{n-1}-3h/2} } w_M(t_1, \ldots , t_l; z_1, \ldots , z_n) 
& = &
 w_{N^{\prime}}(t_1,\ldots , t_{l-1}; z_1, \ldots , z_{n-2}) \times \\
& & {\hspace{-3cm}} \times 
\displaystyle{\prod_{k=1}^{n-2} { { z_{n-1}- z_k -h} 
   \over {z_{n-1} -z_k } } \prod_{a=1}^{l-1} ( t_a - z_{n-1} + h/2 ),} 
\eeqcol
\label{res-form-4}
\eeq

which is obviously independent of the pole structure.

\bigskip

{\bf Lemma 2:} {\it The form factor (\ref{sol-1}) has a KR pole 
$z_n = z_{n-1} -h$ if $n-1 \in M$ and $n \nin M$ and 
arises from the singular hyperplanes $t_a = z_{n-1} -3 h/2$. 

The form factor (\ref{sol-1}) has a KR pole 
$z_n = z_{n-1}  + h$ if $n-1 \nin M$ and $n \in M$ and 
arises from the singular hyperplanes $t_a = z_{n} -3 h/2$.

There are no KR poles if both $n-1, n \in M$ or $n-1, n \nin M$.
}

\medskip

{\sc Proof:} In (\ref{poles}) we have collected the singular hyperplanes 
of the integral (\ref{qKZ-sol}). We could think of evaluating 
the integrals by means of the Leray residue method. Having 
done this one realises that the source for poles of the 
kind $ z_n = z_{n-1} -h$ or $z_n = z_{n-1}  + h$ can only 
reside in expressions arising from the phase function 
(\ref{phi-def}). Now a kinematical pole of the first type 
appears if we take  $t_a = z_{n-1} -3 h/2$. This singularity 
shows up iff it is present in $w_M$ (\ref{w-def}) and 
hence if $n-1 \in M$. We have to check that if this 
is the case the presence of $n$ in $M$ leads to a 
vanishing contribution. This is easy to see because e.g.

\beq
 {\rm{res}}_{t_l= z_{n-1}-3 h/2} w_M |_{z_n=z_{n-1}-h} =0.   
\eeq

The proof of the second statement is similar. The last statement 
is then a consequence of (\ref{res-form-1}) and the first two 
statements in Lemma 2. 

\bigskip

We can now check under which conditions of $P$ the form factor 
satisfies the kinematical residue equations (\ref{KR1}) and 
(\ref{KR2}) in the thermalised case. 

First of all it is clear from the Proposition that if both 
$n-1,n \in M$ or $n-1, n \nin M$ the residue of $F_M$ at 
the points in question vanishes. This means that in this 
case $F_M$ satisfies the equations (\ref{KR1}) and (\ref{KR2}) 
since the charge conjugation matrix $C$ vanishes here. 

Next, we mention that due to (\ref{poles}) 
for the other cases discussed in 
the Proposition, {\it i.e.} $(\eps_{n-1},\eps_n) = (-, +)$ 
or $=(+, -)$ we do not encounter the pinching of integration 
contours which occurs in the case of standard form 
factors in the model \cite{SMIR2}.

\medskip

The verification  of the residue equation (\ref{KR1}) for 
$z_n= z_{n-1} -h$ is more or less straightforward. 
We use the proposition of this section and evaluate 
the multiple integral ${\cal{I}}^P_M$ at 
$t_l= z_{n-1} - 3 h/2$. This pole is present for 
$(\eps_{n-1},\eps_n) = (-, +)$ and for 
$(\eps_{n-1},\eps_n) = (+, -)$. Then we take the 
residue of the remaining expression at the point in 
question and use the identities (\ref{phi-psi-id}) and 
(\ref{minform-id}). 

Let us state that we set $\delta =0$ if 
$(\eps_{n-1},\eps_n) = (-, +)$ and 
$\delta=1$ if $(\eps_{n-1},\eps_n) = (+, -)$, and 
that we abbreviate the parameter $P$ in 
(\ref{sol-1}) by $P_{l,n}$. The result of 
the procedure outlined before is then

\beq
\beqcol
\displaystyle{
\res_{z_n=z_{n-1} -h } F_M(z_1, \ldots , z_n) } & = & 
\displaystyle{
(-1)^{\delta} \prod_{n-2\ge i > j \ge 1} f(z_i-z_j) \;
{  { c_{l-1}} \over {(2 \pi i)^{l-1}}} \times }      \\
&  & {\hspace{-4cm}} \times 
\displaystyle{ 
\int_{C_1}{\rm{d}}t_1 \cdots \int_{C_{l-1}}{\rm{d}}t_{l-1}
w_N(t_1,\ldots ,t_{l-1}; z_1, \ldots , z_{n-2} ) 
P_{l,n}|_{ {t_l=z_{n-1}-3 h/2}\atop{z_n=z_{n-1}-h}} } \\
& & {\hspace{-4cm}} \times \displaystyle{
\prod_{a=1}^{l-1} \prod_{j=1}^{n-2}\Phi (t_a-z_j) 
\prod_{a<b}^{l-1} \psi(t_a-t_b)  \;
\prod_{j=1}^{n-2} e^{-i \pi (z_{n-1} - z_j - 3 h/2) / p } } \times \\
& & {\hspace{-4cm}} \times 
\displaystyle{ \prod_{a=1}^{l-1} 
{  { e^{-2 \pi i ( t_a-z_{n-1}+h/2 ) /p }} \over
  {\sin { {\pi}\over{p}} (t_a-z_{n-1}+ h/2) 
  \sin { {\pi}\over{p}} (t_a-z_{n-1}+ 5 h/2) }} .}  
\eeqcol
\label{kinres-1}
\eeq

This in turn means that $F_M$ satsifies the first kind of 
kinematical residue equations if 

\beq
P_{l,n} \displaystyle{ \prod_{a=1}^{l-1} 
{  { e^{-2 \pi i ( t_a-z_{n-1}+h/2 ) /p }} \over
  {\sin { {\pi}\over{p}} (t_a-z_{n-1}+ h/2) 
  \sin { {\pi}\over{p}} (t_a-z_{n-1}+ 5 h/2) }} 
\prod_{j=1}^{n-2} e^{-i\pi (z_{n-1}-z_j-3 h/2)/p}}  
\sim P_{l-1,n-2}.
\label{Prec-1}
\eeq

The ``$\sim$'' means that this is an equality up to 
a phase, which is permitted by the requirements on 
the form factors.

\bigskip

To check the second type of kinematical poles at 
$z_{n}=z_{n-1} + h$, which is at least not explicitly 
necessary in the level zero case, we take a different 
strategy. Rather than to take the residue directly at 
this point we perform an analytic continuation and 
evaluate the residue at $z_n=z_{n-1} +p -h$. The fact that 
the corresponding residue equations are equivalent is 
shown in cite \cite{MAX1}. In addition it may be nice 
to see how the $S$-matrix factors in (\ref{KR2}) 
come about. 

We use the property (\ref{qKZ-proof-1}) which was used 
in the construction of solutions of the qKZ equations 
in section 3. Up to total derivatives we have

\beq
w_{\eps_1, \ldots , \eps_{n-1},\eps_n}
(\cdot ; z_1, \ldots , z_{n-1},z_n ) 
\sim 
Z_{\hat{1}} w_{\eps_n, \eps_1, \ldots , \eps_{n-1}}
(\cdot ;z_n, z_1, \ldots , z_{n-1} ). 
\label{a-c-1}
\eeq   

Here $Z_{\hat{1}}$ acts on the first index $\eps_n$ 
associated with the variable $z_n$. 

Now we move $z_n$ back to the place $n-1$ using the 
$R$-matrix property of the rational hypergeometric 
basis functions according to (\ref{qKZ-proof-2}). 
Hence we get up to total derivatives

\beq
\beqcol
w_{\eps_1, \ldots , \eps_{n-1},\eps_n}
(\cdot ; z_1, \ldots , z_{n-1},z_n ) 
& \sim &  \phi({\bf{t}}, {\bf{z}} ) \;
R^{\eps^{\prime}_1 \alpha_1}_{\eps_1 \eps_n}(z_1-z_n-p) 
R^{\eps^{\prime}_2 \alpha_2}_{\eps_2 \alpha_1}(z_2-z_n-p)
\cdots \\
& & {\hspace{-5cm}} \cdots
R^{\eps^{\prime}_{n-2} \alpha_{n-2}}_{\eps_{n-2} \alpha_{n-3}}
(z_{n-2}-z_n-p)  
w_{\eps_1^{\prime}, \ldots , \eps^{\prime}_{n-2},
  \alpha_{n-2},\eps_{n-1}} (\cdot ; z_1, z_2,\ldots, z_n+p, z_{n-1} ).
\eeqcol
\label{a-c-2}
\eeq

Hence we can now evaluate the integral ${\cal{I}}_M^P$ with 
this replacement in mind at $t_l=z_n+p -3 h/2$ 
much in the same way as above. 

The result then is

\beq
\beqcol
\displaystyle{
\res_{z_n=z_{n-1} +h-p } F_M(z_1, \ldots , z_n) } & = & 
\displaystyle{
 (-1)^{\delta+1} \prod_{n-2\ge i > j \ge 1} f(z_i-z_j) \;
{  { c_{l-1}} \over {(2 \pi i)^{l-1}}} \times 
\prod_{i=1}^{n-2} S(z_{n-1}-z_i) }      \\
&  & {\hspace{-4cm}} \times 
\displaystyle{ 
\int_{C_1}{\rm{d}}t_1 \cdots \int_{C_{l-1}}{\rm{d}}t_{l-1}
w_N(t_1,\ldots ,t_{l-1}; z_1, \ldots , z_{n-2} ) 
P_{l,n}|_{ {t_l=z_{n}+p-3 h/2}\atop{z_n=z_{n-1}+h-p}} } \\
& & {\hspace{-4cm}} \times \displaystyle{
\prod_{a=1}^{l-1} \prod_{j=1}^{n-2}\Phi (t_a-z_j) 
\prod_{a<b}^{l-1} \psi(t_a-t_b)  \;
\prod_{j=1}^{n-2} e^{-i \pi (z_{n-1} - z_j -  h/2) / p } } \times \\
& & {\hspace{-4cm}} \times 
\displaystyle{ \prod_{a=1}^{l-1} 
{  { e^{-2 \pi i ( t_a-z_{n-1} - h/2 ) /p }} \over
  {\sin { {\pi}\over{p}} (t_a-z_{n-1}- h/2) 
  \sin { {\pi}\over{p}} (t_a-z_{n-1}+ 3 h/2) }} .}  
\eeqcol
\label{kinres-2}
\eeq

Hence the kinematical residue equation (\ref{KR2}) is 
satisfied if

\beq
P_{l,n} \displaystyle{ \prod_{a=1}^{l-1} 
{  { e^{-2 \pi i ( t_a-z_{n-1} - h/2 ) /p }} \over
  {\sin { {\pi}\over{p}} (t_a-z_{n-1}- h/2) 
  \sin { {\pi}\over{p}} (t_a-z_{n-1}+ 3 h/2) }} 
\prod_{j=1}^{n-2} e^{-i \pi (z_{n-1}-z-j-h/2)/p} }  
\sim P_{l-1,n-2}.
\label{Prec-2}
\eeq

We are going to discuss some important solutions of the 
equations (\ref{Prec-1}) and (\ref{Prec-2}) in the 
following section.

Obviously, the value of $n/2-l$ an invariant under 
the kinematical residue equations. This invariant 
can be interpreted as the isospin of form factor 
sequence. This means that by (\ref{Prec-1}) and 
(\ref{Prec-2}) we get sequences of solutions of 
all form factor equations for each value of 
the isospin. 

Let us collect the main results in a Theorem. 
\bigskip

{\bf Theorem 4}: {\it There exist infinite sequences  

\beq
\label{th2}
\displaystyle{
F_M(z_1, \ldots , z_n) = { {c_l}\over { (2\pi i)^l}} 
\prod_{n \ge i > j \ge 1} f(z_i - z_j) \times
{\cal{I}}^{P_{l,n}}_M(z_1, \ldots , z_n),} 
\nonumber
\eeq

of solutions of the qKZ-equation with the $R$-matrix being 
the $S$-matrix (\ref{S-matrix}) of the Thirring model. 
Each sequence is labelled by the invariant $n/2-l$. 
If in addition $P_{l,n}$ satisfies the recursion relations 
(\ref{Prec-1}) and (\ref{Prec-2}) then the members 
of each sequence are linked by (\ref{KR1}) and hence provide 
replica-deformed form factors of local operators in 
the Thirring model with isospin $n/2-l$.}
\bigskip

Of course every sequence (\ref{th2}) can be multiplied
pointwise with a sequence $Q(z_1,\ldots,z_n)$ without changing
the properties described, provided $Q(z_1,\ldots,z_n)$ is 
completely symmetric and $i\beta$-periodic in all arguments,
and satisfies

\beq
Q(z_1,\ldots,z_n)\bigg|_{z_n = z_{n-1} \pm i\pi} = 
Q(z_1,\ldots,z_{n-2})\,.
\label{chargeEV}
\eeq

Alternatively such sequences arise as ambiguities in the solution
of the recursive equations for the polynomials $P_{l,n}$.
Physically the solutions of (\ref{chargeEV}) correspond to 
eigenvalues of a local conserved charge. As shown in \cite{MAX1}
the structure of these eigenvalues for $\beta \neq 2\pi$ is quite 
different from that in the undeformed case. Also the eigenvalues 
in both cases are not automatically in one-to-one correspondence, 
though they
can be made so by imposing suitable minimality conditions. 
In any case, modulo their respective ambiguities, the completeness
result for section 5 (together with its undeformed counterpart) 
now implies the announced correspondece (\ref{qcorr}): 
The form factor sequences in the replica-deformed model are 
essentially in one-to-one correspondence to their undeformed QFT 
counterparts. In particular this allows one to identify the 
replica copy of the form factors of a local operators.  
As an illustration we present the replica-deformed form factors
of the Noether current in the next section.

\bigskip
\bigskip

\section{Deformed form factors of the Noether current}

\bigskip
\bigskip

In this section we present solutions to the equations 
(\ref{Prec-1}) and (\ref{Prec-2}). As it was mentioned 
in the introduction, the form factor approach enables 
to classify the full space of local operators of a given 
model. This arises from the fact that there exist kernel 
solutions to the equations (\ref{Prec-1}) and (\ref{Prec-2}). 
This feature is illustrated in \cite{SMIR1} where the 
local fields in the sine-Gordon model have been 
``counted'', and in \cite{MAP2} where the full space 
of local operators in the Sinh-Gordon model has been 
found. In the latter model the local operators in 
one superselection sector were identified with the 
integer powers of the elementary field appearing in 
the Lagrangian together with an exponential field. 

It should be possible to mimic the analysis of the 
local fields of \cite{SMIR3} in the present case as 
well. However, we would like to present in this 
section solutions to (\ref{Prec-1}) and (\ref{Prec-2}) 
which in the limit $\beta = 2 \pi$ turn into the form factors 
corresponding to the $SU(2)$ 
Noether current in the Thirring model. These form factors are 
well understood in the standard case and it has been 
shown in \cite{NPT} how these form factors arise 
from the qKZ-solutions at level zero and how they 
come about from the trace formulae of vertex operators 
in the Yangian double of \cite{KLP}.

\medskip

Let $j_{\mu}^a$ be the Noether current associatted with the 
Lagrangian (\ref{lagrange}); $\mu$ is the Lorentz index and $a$ the 
$SU(2)$-index as introduced in section 3. It is convenient to 
switch to lightcone coordinates both in Minkowski space and in 
internal space, i.e.

\beq
j_{\pm}^a = j_0^{a} \pm j_1^a\,,\qquad 
j_{\sigma}^{\pm} = j_{\sigma}^1 \pm i \; j_{\sigma}^2 , \qquad
\sigma = \pm.
\label{current}
\eeq

Then we can introduce an index $\tau=\pm, 3$ labelling the 
components of the lightcone currents as $j^{\tau}_{\sigma}$.
A solution of the equations (\ref{Prec-1}) and (\ref{Prec-2}) 
is given by

\beq
P^{\sigma}_{l,n} = d^{\sigma}_{l,n} 
\displaystyle{ e^{-i\pi(2 l -n/2+\sigma)\sum_{j=1}^{n}z_j \;/p}
\prod_{a=1}^{l} e^{2\pi i (l+\sigma)t_a /p} 
\prod_{1 \le a < b \le l} \left( \sin { {\pi}\over{p}} (t_a-t_b-h)
                          \sin { {\pi}\over{p}} (t_a-t_b+h) \right)},
\label{P-sol-1}
\eeq

where the constant $d^{\sigma}_{l,n}$ is 
\beq
\displaystyle{
d^{\sigma}_{l,n} = \exp \left( -{ { i \pi }\over{p}}
( n-2l -2 (\sigma +1) ) \; l \right) }.
\label{dln}
\eeq

Comparing this with the level zero counterpart \cite{SMIR2,NPT} 
we find that this choice of $P_{l,n}$ corresponds to the form factors 
of the current $j^{-}_{\sigma}$. The form factors of the other 
components of $j_{\sigma}^{\tau}$ can obviously be obtained 
by applying the operator $\Sigma^{+}$ once or twice to the 
corresponding form factor.

A closer look on the solution (\ref{P-sol-1}) shows that 
the corresponding form factor does satisfy the first 
two of the form factor equations (\ref{watson}) with a 
factor $\eta \ne 1$. But still we have $| \eta| = 1$ as 
it should be. The reasons why this feature appears is 
disussed in detail in \cite{SMIR2}. 

\medskip

Of course, in order to establish that these solutions do 
really correspond to the replica-deformed form factors of 
the currents we have to prove several conditions like 
in the level zero case, see section 6 of \cite{SMIR2}. 
The most important of these conditions is the current 
conservation condition $\partial_{\mu} \; j_{\mu}^a =0$. 

To verify this in the deformed framework involves the 
knowledge of the deformed eigenvalues of the lightcone momenta. 
We remind that the kinematics in the deformed model 
is quite different from the usual model as was outlined 
in section 2. Even 
though these eigenvalues have been characterised and 
computed for some low values of the particle number 
$n$ in \cite{MAX1}, a general formula is not known at present. This 
certainly deserves a further study also in the light 
of the characterisation of other operators in the 
replica-copy of the Thirring model as well. 

Therefore, we can at present only conjecture that the 
form factors characterised by the solutions (\ref{P-sol-1}) 
do correspond to the Noether-current operators in 
the replica-copy of the Thirring model. Nevertheless 
we can give an argument to support our conjecture. 
Let $P_{\pm}$ be the eigenvalues of the lightcone 
momenta on an $n$-particle state in the $\beta \ne 2 \pi$ 
copy of the Thirring 
model according to \cite{MAX1}. We denote by 
$\Psi_{\sigma}(z_1, \ldots , z_n )$ the functions 
which we conjecture to be the form factors of the 
currents $j^{-}_{\sigma}$. 

The explicit difference of $\Psi_{+}$ and $\Psi_{-}$ 
resides first in the purely $z$-dependent part of 
(\ref{P-sol-1}). This bit is not affected by the 
integration. The second $\sigma$-dependent part 
comes from the first product term in (\ref{P-sol-1}). 
This term is trivial in the sense that it responsible 
neither for poles nor for zeros of the integrand. In 
addition it is $p$-periodic. Therefore it is clear 
that there is a scalar function $R(z_1,\ldots ,z_n)$ 
such that for any particle number we have an 
identity $\Psi_{+} = R \Psi_{-}$. According to what 
was said at the end of the previous section we can now 
redefine for any number of arguments $n$ 
the form factors of the currents by setting

\beq
J_{+} := P_{+} R^{-1/2} \Psi_{+}, 
\qquad
J_{-} := -P_{-} R^{1/2} \Psi_{-}.
\eeq

Then we can write the current conservation as follows:

\beq
P_+ J_-  +  P_- J_+ = P_+ P_- R^{-1/2} \left( \Psi_+ - R \Psi_- \right),
\eeq

which is zero due to the aforementioned identity.

\bigskip
\bigskip

\section{Discussion}

\bigskip
\bigskip
 
In this paper we have been trying to illustrate the natural 
connection between rational solutions of the qKZ equation and 
the replica-deformed form factors in the $SU(2)$-invariant 
Thirring model. Here we took up two independent directions 
of recent research. 

The first one is Niedermaier's formalism showing that every 
integrable massive quantum field theory in $1+1$-dimensions admits an 
$S$-matrix preserving deformation, which can be 
interpreted as providing a ``replica-copy'' of the original 
QFT with an off-critical Unruh temperature. 
It was shown in \cite{MAX1} that this replica-copy can be 
described exactly in terms of its form factors, the latter 
being solutions of a a modification of the standard form 
factor equations \cite{KW,SMIR1}. The present paper is the first 
in which these equations have been solved for a particular model 
in full generality. 

As a technical tool we took advantage of another recent development, 
namely the study of the solutions of the rational $sl_2$ qKZ equation 
\cite{TV,MV}. As one could suspect by looking at the first set of 
modified form factor equations (\ref{watson}), their 
solution in the Thirring model case should somehow 
reside in the space of solutions of the qKZ-equation 
for a particular representation of $sl_2$. The 
question then is whether among these solutions 
we can find some satisfying the deformed kinematical 
residue equations (\ref{KR1}) as well.  

The main purpose of this work was to show that this is 
indeed possible. In outline we rewrote the qKZ-solutions of \cite{TV,MV} 
at generic level in a way faciliating the comparison with the standard 
level zero situation. 
We argued that this form of the solution naturally 
shares the property that the algebraic Bethe vectors 
remain unchanged under the replica thermalisation. This 
fact could have been expected since the modified 
form factor equations use the usual scattering matirx. 
We then supplemented our qKZ-solutions with the 
minimal form factor to define an object (\ref{sol-1}) 
having only the polynomial $P$ unspecified. 
Using the pole structure of the qKZ-solutions we derived 
conditions on $P$ from the modified kinematical residue 
equations and solved these conditions to conjecture the 
deformed form factors of the Noether-current in the $SU(2)$ 
Thirring model.  

\medskip

To summarise we have shown that the Unruh-thermalised form 
factors arise naturally from qKZ-solutions at generic 
level as the standard form factors do at level zero.

\medskip
 
We may also anticipate thath the $\beta \rightarrow 0$ limit of our 
construction bears an unexpected relation to dimensionally 
reduced quantum gravity \cite{MH}.  

\medskip

It should now be possible to derive integral representations 
of the form factors for the thermalised sine-Gordon model 
using the results on trigonometric qKZ-solutions as 
derived in \cite{TV2}. It would be particularly 
interesting to construct the thermalised form factors 
of the breather sectors. This would involve also 
the construction of modified fusing relations for 
the form factors. The latter haven't been supplied 
in \cite{MAX1}. In addition solving this problem would 
also naturally lead to an integral representation of 
the thermalised form factors in the Sinh-Gordon model, 
a problem which has only partially been solved in \cite{MAX1}.

\bigskip
\bigskip
\bigskip

{\bf Acknowledgement:}

I am grateful to M.~Niedermaier for many discussions on the 
replica-deformed form factor equations and to M.~Flohr for 
conversations on hypergeometric integrals. 
Thanks go also to A.~Nakayashiki for a useful correspondence 
on the work \cite{NPT}. 

This work was supported by EPSRC (grant GR/L26216).

\end{document}